\documentclass[pdflatex,sn-mathphys-num]{sn-jnl}

\usepackage{graphicx}%
\usepackage{multirow}%
\usepackage{amsmath,amssymb,amsfonts}%
\usepackage{amsthm}%
\usepackage{mathrsfs}%
\usepackage[title]{appendix}%
\usepackage{xcolor}%
\usepackage{textcomp}%
\usepackage{manyfoot}%
\usepackage{booktabs}%
\usepackage{algorithm}%
\usepackage{algorithmicx}%
\usepackage{algpseudocode}%
\usepackage{listings}%

\theoremstyle{thmstyleone}%
\theoremstyle{thmstyletwo}%

\theoremstyle{thmstylethree}%

\begin{document}

\title[Article Title]{A new solution of Einstein-Maxwell field equations in isotropic coordinates
}


\author[1]{\fnm{B. S.} \sur{Ratanpal}}\email{bharatratanpal@gmail.com}

\author*[1]{\fnm{Bhavesh} \sur{Suthar}}\email{bhaveshsuthar.math@gmail.com}

\affil[1]{\orgdiv{Department of Applied Mathematics, Faculty of Technology \& Engineering}, \orgname{The Maharaja Sayajirao University of Baroda}, \orgaddress{\city{Vadodara}, \postcode{390001}, \state{Gujarat}, \country{India}}}



\abstract{In this work, we provide a novel exact solution to the Einstein-Maxwell field equations in isotropic coordinates for charged anisotropic matter distribution, achieved by considering one of the metric potentials as well as a particular form of pressure anisotropy and electric field. We analysed all physical features of a realistic star to ensure the model's viability, and we observed that it met all the physical plausibility conditions. The gravitational potentials and matter variables are well-behaved and regular throughout the distribution. All energy conditions have been verified using graphical representations. The stability criteria of the model have been discussed via different conditions. A detailed physical analysis of the star \textbf{4U 1538-52} demonstrates that our model is viable.}

\keywords{General relativity, Exact solutions, Isotropic Coordinates, Anisotropy}



\maketitle

\section{Introduction}\label{sec1}
The pursuit of exact solutions to Einstein's field equations has led to in a diverse set of models representing relativistic compact objects. Schwarzschild \cite{Schwarzschild} discovered the solution to Einstein's field equation, which explains the neighbourhood of compact objects. Delgaty and Lake \cite{DL1998} described the physical model of a compact star by obtaining exact solutions to Einstein's field equations. Numerous analytical solutions were developed by them, each of which describes a perfect static fluid satisfying all the requirements for physical acceptability. These exact solutions have also allowed us to investigate cosmic censorship and the emergence of naked singularities \cite{Pjoshi}. Giuliani and Rothman \cite{giuliani} \& Bohmer and Harko \cite{bohmer} used specific models to determine the absolute stability limit for charged spheres. Mak and Harko \cite{Mak2002}, Komathiraj and Maharaj \cite{komathiraj2007}, and Thirukkanesh and Maharaj \cite{TM2008} explored strange stars using various models of relativistic matter.

In 1998, Delgaty and Lake \cite{DL1998} observed that out of 127 static, spherically symmetric solutions to Einstein's field equations, only a few of them in isotropic coordinates were considered physically relevant for modeling compact stellar objects. Nariai \cite{Nariai} demonstrated that any line element given in a canonical, such as the Schwarzschild coordinate system, can be simply transformed into the isotropic coordinate system. However, vice versa is not always feasible. Each of the three spatial dimensions is treated equally in the isotropic line elements. Isotropic coordinates may bring novel perspectives and perhaps leading to new solutions. Einstein's field equations have been simplified by Mak and Harko \cite{MH2005} to two independent Riccati differential equations, for which a matrix transformation in isotropic coordinates yields three classes of solutions. The exact solutions to Einstein's field equation for a spherically symmetric perfect fluid distribution in an isotropic coordinates had been found by Rahman and Visser \cite{RV2002} and Lake \cite{Lake2003}. Many researchers have obtained exact solutions to Einstein's field equations in an isotropic coordinate system. Noteworthy contributions in this context include the works of Ngubelanga and Maharaj \cite{NM2013}, Newton et al. \cite{Newton2014}, Pant et al. \cite{Pant2015}, Ngubelanga et al. \cite{N2015}, Govender and Thirukkanesh \cite{Govender2015}, Bhar et al. \cite{Bhar2018}, and Ratanpal and Suthar \cite{Suthar2024}.

At ultrahigh densities, the radial as well as transverse pressures inside the stellar interior might be unequal, and the pressure within the stellar object may be anisotropic \cite{Ruderman1972}, which may occur for a variety of causes \cite{BL1974}. The effects of local anisotropy on astrophysical objects and their causes have been studied by a number of authors \cite{HS1997}, \cite{HS1997-1}, \cite{MH2002}, \cite{MH2002-1}, \cite{Chan2003}, and \cite{HM2004}. Letelier \cite{L1980} postulated that pressure anisotropy may be caused by a mixing of two gases, such as ionised hydrogen and electrons. According to Weber \cite{Weber}, high magnetic fields can result in an anisotropic pressure component inside a compact sphere. Herrera and Santos \cite{HS1995} investigated the impact of slow rotation on stars. Dev and Gleiser \cite{DG2002} shown how pressure anisotropy affects the physical characteristics, structure, and stability of compact stars. Usov \cite{U2004} proposed that anisotropy may also be formed when a net electric charge is present.

Ivanov \cite{I2002} demonstrated for the first time that the presence of charge in a perfect fluid inhibits the growth of spacetime curvature, hence avoiding singularities. Dev and Gleiser \cite{DG2003} and Gleiser and Dev \cite{GD2004} demonstrated that the existence of anisotropic pressures in charged matter increases the stability of the configuration under radial adiabatic perturbations as compared to isotropic matter. By adding charge, the critical radius for instability may be reduced. The physical characteristics of charged spheres can be described by finding the exact solutions of the Einstein-Maxwell equations for gravitational fields with spherical symmetry. This system has developed numerous applications in stellar systems \cite{KM2007}, \cite{Sharma2001}. There are several methods for solving the Einstein-Maxwell system in the literature, such as those proposed by Sharma et al. \cite{Sharma2001}, Feroze and Siddiqui \cite{FS2011}, Feroze and Siddiqui \cite{FS2014}, Ratanpal et al. \cite{Ratanpal2015}, Ratanpal et al. \cite{ratanpal2017}, Thirukkanesh et al. \cite{thirukkanesh2022}, and Mathias et al. \cite{mathias2023}. Einstein-Maxwell field equations seem to be very important for describing ultracompact objects.

In the present article, the exact solution to the Einstein-Maxwell field equations in an isotropic coordinate system is obtained by considering a particular ansatz for metric potential and a certain kind of pressure anisotropy. It is found that the solution that describes the interior of the star is non-singular and satisfies all physical plausibility conditions. We examined the results for star 4U 1538-52. The article is structured as follows: Section \ref{sec2} describes the spherically symmetric interior spacetime metric and the Einstein-Maxwell field equations for the charged anisotropic distribution of matter in isotropic coordinates. By assuming a physically reasonable gravitational potential as well as a particular form of pressure anisotropy and electric charge, the interior solution is obtained in section \ref{sec3}. Section \ref{sec4} describes the junction conditions required for the smooth matching of the interior spacetime to the R-N exterior spacetime. This section also addressed the solution of the Einstein-Maxwell field equations. Section \ref{sec5} illustrates physically acceptable conditions for anisotropic models. Finally, in Section \ref{sec6}, we conclude the work.

\section{Einstein-Maxwell field equations}\label{sec2}
The interior of an anisotropic fluid sphere in an isotropic coordinate system is expressed by the static and spherically symmetric line element
\begin{equation} \label{IMetric}
    ds^{2}=-A^2(r)dt^2+B^2(r)\left[dr^2+r^{2}\left(d\theta^{2}+\sin^{2}\theta d\phi^{2} \right)\right],
\end{equation}
where $A(r)$ and $B(r)$ are arbitrary functions of the radial coordinate.\\
The energy-momentum tensor for an anisotropic charged matter distribution is
\begin{equation} \label{EMTensor}
    T_{ij}= diag(-\rho-\frac{1}{2}E^{2},p_{r}-\frac{1}{2}E^{2},p_{t}+\frac{1}{2}E^{2},p_{t}+\frac{1}{2}E^{2}),
\end{equation}
where $\rho$, $p_{r}$, $p_{t}$ and $E$ represent the energy density, radial pressure, tangential pressure and electric field intensity, respectively. $\Delta = p_{t}-p_{r}$ is a measure of anisotropy that is repulsive when $p_{t}>p_{r}$ and attractive when $p_{t}<p_{r}$. Radial and tangential pressures are assessed in relation to the comoving fluid four-velocity
\begin{equation} \label{UFV}
    u^{a}=\frac 1{A} \delta^{a}.
\end{equation}
The system of Einstein-Maxwell field equations corresponds to the line element (\ref{IMetric}) and the energy-momentum tensor (\ref{EMTensor}) are
\begin{equation} \label{Rho1}
    8\pi\rho=\frac{-1}{B^{2}}\Bigg[2\frac{B^{''}}{B}-\frac{{B^{'}}^{2}}{{B}^{2}}+\frac{4}{r}\frac{B^{'}}{B}\Bigg]-\frac{1}{2}E^{2},
\end{equation}
\begin{equation} \label{Pr1}
    8\pi p_{r}=\frac{1}{B^2}\Bigg[\frac{{B^{'}}^{2}}{B^{2}}+2\frac{A^{'}}{A}\frac{B^{'}}{B}+\frac{2}{r}\Bigg(\frac{A^{'}}{A}+\frac{B^{'}}{B}\Bigg)\Bigg]+\frac{1}{2}E^{2},
\end{equation}
and
\begin{equation} \label{Pt1}
    8\pi p_{t}=\frac{1}{B^2}\Bigg[\frac{A^{''}}{A}+\frac{B^{''}}{B}-\frac{{B^{'}}^{2}}{B^2}+\frac{1}{r}\Bigg(\frac{A^{'}}{A}+\frac{B^{'}}{B}\Bigg)\Bigg]-\frac{1}{2}E^{2},
\end{equation}
where primes $(')$ represent differentiation with respect to the radial coordinate $r$.
In generating the solution to the above system of equations, we have used geometrised units where the coupling constant and speed of light are taken to be unity. Making use of equations (\ref{Pr1}) and (\ref{Pt1}), we get the differential equation
\begin{equation} \label{Anisotorpy1}
    \Delta=\frac{1}{B^{2}}\Bigg[\frac{A^{''}}{A}+\frac{B^{''}}{B}-2\frac{{B^{'}}^{2}}{B^{2}}-2\frac{A^{'}}{A}\frac{B^{'}}{B}-\frac{1}{r}\Bigg(\frac{A^{'}}{A}+\frac{B^{'}}{B}\Bigg)\Bigg]-E^{2}.
\end{equation}

\section{Choosing potential} \label{sec3}
To develop a physically acceptable model of the stellar configuration, various choices can be made for the metric potential $B(r)$. Following Govender and Thirukkanesh \cite{GT2015}, we choose $B(r)$ as
\begin{equation} \label{B}
    B(r)=\frac{a}{\sqrt{1+br^{2}}},
\end{equation}
where a and b are constants. The gravitational potential $B(r)$ is well-behaved and regular at the origin. It is important to realise that the above-chosen $B(r)$ is physically reasonable for a realistic stellar model. Thirukkanesh et al. \cite{Thirukkanesh2015}, Bhar et al. \cite{Bhar2018}, and Ratanpal and Suthar \cite{Suthar2024} have used the same form of $B(r)$ in isotropic coordinates. With this choice of $B(r)$, the measure of anisotropy $(\Delta)$ reduces to
\begin{equation} \label{Deltaeqn}
    \Delta=\frac{{A}^{''}r{\left(1+br^{2}\right)}^{2}+{A}^{'}\left(1+br^{2}\right)\left(-1+br^{2}\right)+Ab^{2}r^{3}}{Aa^{2}r\left(1+br^{2}\right)}-E^{2}.
\end{equation}
Rearranging equation (\ref{Deltaeqn}) gives
\begin{equation} \label{Eq10}
    A^{''}+\frac{A^{'}\left(br^{2}-1\right)}{r\left(1+br^{2}\right)}+\frac{\left[b^{2}r^{2}-a^{2}\left(1+br^{2}\right)\left(\Delta+E^{2}\right)\right]}{{\left(1+br^{2}\right)}^{2}}A=0.
\end{equation}
Our aim is to generate exact solutions to the Einstein-Maxwell system of equations. To close the system of equations, we choose the expression of anisotropy $(\Delta)$ and electric field intensity $(E^{2})$ as
\begin{equation}\label{Anisotropy2}
     \Delta=\frac{b^{2}r^{2}}{2a^{2}(1+br^{2})},
\end{equation}
\begin{equation}\label{Electric}
     E^{2}=\frac{b^{2}r^{2}}{2a^{2}(1+br^{2})}.
\end{equation}
The motivation for making the particular choice of the anisotropy $(\Delta)$ and electric field $(E^{2})$ is that they make the model physically significant and well-behaved. It turns out $\Delta$ and $E$ remains regular and positive throughout the distribution. It can be seen that $\Delta=0$ and $E=0$ at the center. Substituting (\ref{Anisotropy2}) and  (\ref{Electric}) in (\ref{Eq10}), we obtain
\begin{equation}\label{Eq15}
    {A}^{''}+\frac{{A}^{'}\left(br^{2}-1\right)}{r\left(1+br^{2}\right)}=0.
\end{equation}
Solving (\ref{Eq15}) yields another metric coefficient
\begin{equation} \label{A1}
    A=\frac{C\log(1+br^{2})+2bD}{2b},
\end{equation}
where $C$ and $D$ are constants of integration. Subsequently, the radial pressure ($p_{r}$) and tangential pressure ($p_{t}$) are obtained as
\begin{equation}
    8\pi p_{r}=-\frac{b\left[-4C+2bD\left(2+br^{2}\right)+C\left(2+br^{2}\right)\log\left(1+br^{2}\right)\right]}{a^{2}\left(1+br^{2}\right)\left[2bD+C\log\left(1+br^{2}\right)\right]},
\end{equation}
\begin{equation}
    8\pi p_{t}=-\frac{2b\left[-2C+2bD+C\log(1+br^{2})\right]}{a^{2}(1+br^{2})\left[2bD+C\log(1+br^{2})\right]}.
\end{equation}
Hence, the spacetime metric (\ref{IMetric}) takes the form
\begin{equation} \label{Metric2}
    ds^{2}=-\frac{\big[2bD+C\log(1+b {r}^{2})\big]^{2}}{4b^{2}}dt^{2}+\frac{a^{2}}{(1+br^{2})}\big[dr^{2}+r^{2}{d\theta}^{2}+r^{2}\sin^{2}{\theta}{d\phi}^{2}\big].
\end{equation}

\section{Matching Condition}\label{sec4}
The exterior field of a static spherically symmetric charged fluid distribution is described by the Reissner-Nordström metric
\begin{equation}\label{EMetric}
	ds^{2}=-\left(1-\frac{2M}{r}+\frac{Q^{2}}{r^{2}}\right)dt^{2}+{\left(1-\frac{2M}{r}+\frac{Q^{2}}{r^{2}}\right)}^{-1}dr^{2}+r^{2}\left(d\theta^{2}+\sin^{2}\theta d\phi^{2} \right),
\end{equation}
where $M$ and $Q$ represents the total mass and charge, respectively.
For a physically viable model, the interior spacetime should match the exterior Reissner-Nordström line element throughout the boundary of the star $(r = R)$, with the additional constraint that the radial pressure vanishes at the surface. The junction conditions yield
\begin{equation} \label{C1}
    C=\frac{b\left(8+3 b R^{2}\right) \sqrt{1+b R^{2}}}{8a},
\end{equation}
and
\begin{equation} \label{D1}
    D=\frac{\sqrt{1+b R^{2}} \left[16-\left(8+3bR^{2}\right) \log\left(1+b R^{2}\right)\right]}{16a}.
\end{equation}
Substituting (\ref{C1}) and (\ref{D1}) in (\ref{A1}), we get
\begin{equation} \label{A2}
    A=\frac{\sqrt{1+b R^{2}} \left[16+\left(8+3bR^{2}\right) \log\left(1+b r^{2}\right)-\left(8+3bR^{2}\right) \log\left(1+b R^{2}\right)\right]}{16a}.
\end{equation}
The expression of energy density, radial pressure and tangential pressure then takes the form
\begin{equation} \label{Rho3}
    \rho=\frac{3b\left(8+br^{2}\right)}{4a^{2}\left(1+br^{2}\right)},
\end{equation}
\begin{equation} \label{Pr}
    p_{r}=\frac{b\left[48b\left(R^{2}-r^{2}\right)-k_{1}\log(1+br^{2})+k_{1}\log(1+bR^{2})\right]}{4a^{2}\left(1+br^{2}\right)\left[16+k_{3}\log(1+br^{2})-k_{3}\log(1+bR^{2})\right]},
\end{equation}
and
\begin{equation} \label{Pt}
    p_{t}=\frac{b\left[16b\left(3R^{2}-r^{2}\right)-k_{2}\log(1+br^{2})+k_{2}\log(1+bR^{2})\right]}{4a^{2}\left(1+br^{2}\right)\left[16+k_{3}\log(1+br^{2})-k_{3}\log(1+bR^{2})\right]},
\end{equation}
where
\begin{equation*}
    k_{1}=64+9b^{2}r^{2}R^{2}+24b\left(R^{2}+r^{2}\right),
\end{equation*}
\begin{equation*}
    k_{2}=64+3b^{2}r^{2}R^{2}+8b\left(3R^{2}+r^{2}\right),
\end{equation*}
\begin{equation*}
    k_{3}=8+3bR^{2}.
\end{equation*}

\section{Physical Analysis}\label{sec5}
Kuchowicz \cite{K1972}, Buchdahl \cite{B1979}, and Knutsen \cite{K1988} proposed criteria to ensure the model is physically reasonable and well-behaved. In the following subsections, we examined acceptability and stability under different conditions. For this purpose, we have considered the radius of a compact star proposed by Gangopadhyay et al. \cite{Gangopadhyay}. We analysed the results for the particular star 4U 1538-52 with a radius $R = 7.87 km$. We choose $a = 0.843 km^{-2}$ and $b = 0.006 km^{-2}$. Along with this data, we generated plots of various model parameters and investigated other physical aspects to get a more realistic stellar configuration.

\subsection{Regularity of Gravitational Potential}
The model should be free from physical and geometric singularities. In order to observe this, we need to verify if the gravitational potentials $A(r)$ and $B(r)$ are regular, continuous and finite.\\
In our model, 
\begin{equation*}
    A^{2}\left(0\right)=\frac{\left(1+bR^{2}\right){\left[16-\left(8+3bR^{2}\right) \log\left(1+bR^{2}\right)\right]}^{2}}{256a^{2}}
\end{equation*}
and
\begin{equation*}
    B^{2}\left(0\right)=a^{2}
\end{equation*}
which are constants.\\
Also, we have $\left(A^{2}\left(r\right)\right)^{'}_{r=0}=\left(B^{2}\left(r\right)\right)^{'}_{r=0}=0,$
which demonstrates that the gravitational potentials are regular at the origin.

\subsection{Density and Pressure Profile}
\begin{figure}
\centering
    \includegraphics[height=.25\textheight]{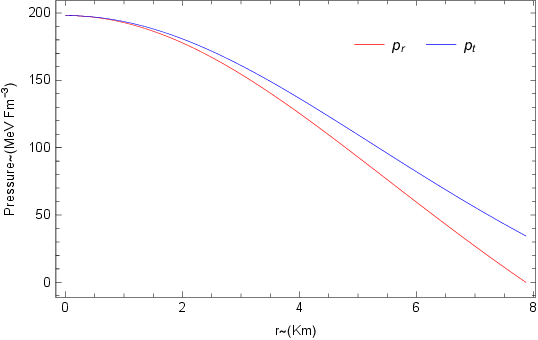}
    \caption{Variation of pressures ($p_{r}$, and $p_{t}$) with respect to radius $(r)$}
    \label{fig:Figure. 2}
\end{figure}

The energy density and radial pressure should be finite and positive inside the stellar interior. Since
\begin{equation*}
\rho\left(0\right) = \frac{6b}{a^{2}}
\end{equation*}
and
\begin{equation*}
    p_{r}\left(0\right) = \frac{12b^{2}R^{2}+2b\left(8+3bR^{2}\right)\log\left(1+bR^{2}\right)}{a^{2}\left[16-\left(8+3bR^{2}\right)\log\left(1+bR^{2}\right)\right]},
\end{equation*}
the energy density is positive and regular at the center if $b > 0$, and the positiveness of radial pressure at the center depends on the choice of parameters a, b and R.
Furthermore, the energy density and pressure should decrease monotonically throughout the distribution. We have
\begin{equation*}
    \frac{d\rho}{dr}=\frac{-21 b^{2}r}{2a^{2}\left(1+br^{2}\right)^{2}}<0,
\end{equation*}
indicating that the energy density decreases with r. Also, the gradients of radial and tangential pressure are negative within the stellar body, as illustrated in Figure \ref{fig:Figure. 2}, ensuring that radial and tangential pressure decrease with the radial parameter.

\begin{figure}
\centering
    \includegraphics[height=.25\textheight]{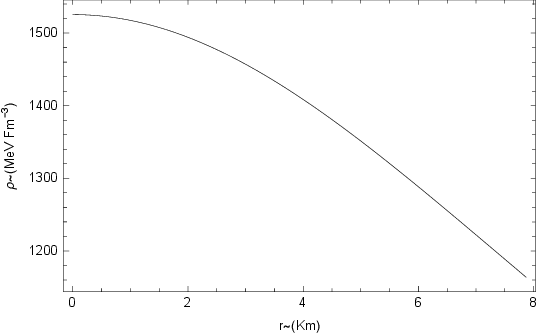}
    \caption{Variation of energy density ($\rho$) with respect to radius $(r)$}
    \label{fig:Figure. 1}
\end{figure}

\subsection{Pressure Anisotropy}
\begin{figure}
\centering
    \includegraphics[height=.25\textheight]{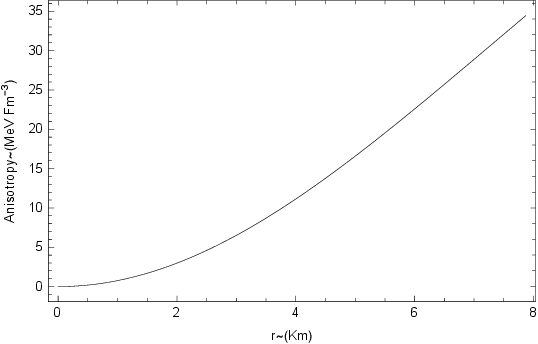}
    \caption{Measure of Anisotropy ($\Delta$) against radial coordinate $(r)$}
    \label{fig:Figure. 3}
\end{figure}

The measure of anisotropy $\left(\Delta = p_{t}-p_{r}\right)$ should be zero at the origin and increasing towards the surface. Since the radial pressure $\left(p_{r}\right)$ is equal to the tangential pressure $\left(p_{t}\right)$ at the origin, the anisotropy $\left(\Delta\right)$ vanishes at the origin, while at the boundary $r=R$ it is positive and is equivalent to the tangential pressure $\left(p_{t}\right)$. The measure of anisotropy $\left(\Delta\right)$ is plotted in Figure \ref{fig:Figure. 3}.

\subsection{Stability Conditions}
\subsubsection{Causality Condition}
\begin{figure}
\centering
    \includegraphics[height=.25\textheight]{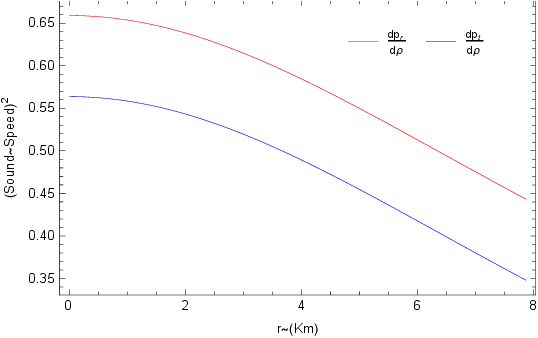}
    \caption{Squared Sound Speed against radial coordinate ($r$)}
    \label{fig:Figure. 4}
\end{figure}

Since the fluid in a compact stellar system is non-exotic, the causality condition needs to be satisfied while determining the velocity of sound in the medium. The causality condition demands that the sound speed should not exceed the speed of light and it must be within the limit in the interior of the star, i.e., the squared radial sound speed $\left({v^{2}_{r}}\right)$ and squared tangential sound speed $\left({v^{2}_{t}}\right)$ should lie between $[0,1]$. The radial speed $\left({v^{2}_{r}}\right)$ and tangential speed  $\left({v^{2}_{t}}\right)$ of sound are given as
\begin{equation*}
    {v^{2}_{r}}=\frac{dp_{r}}{d\rho},    
    {v^{2}_{t}}=\frac{dp_{t}}{d\rho}.
\end{equation*}
Due to the complexity of the expressions, we analysed this condition using a graphical representation. It is clear from the Figure \ref{fig:Figure. 4} that $0\leq{v^{2}_{r}}\leq1$ $\&$ $0\leq{v^{2}_{t}}\leq1$ everywhere within the stellar structure. Therefore, one can say that the causality condition is well satisfied.

\subsubsection{Adiabatic Index}
\begin{figure}
\centering
    \includegraphics[height=.25\textheight]{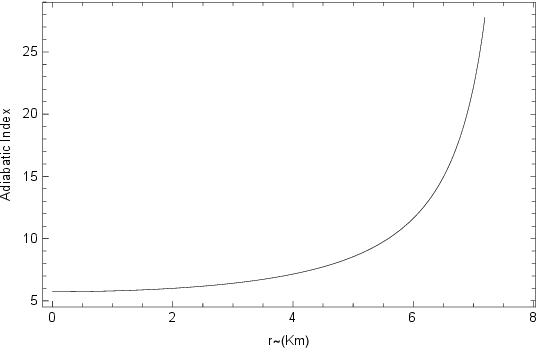}
    \caption{Adiabatic Index ($\Gamma$) against radial coordinate $(r)$}
    \label{fig:Figure. 5}
\end{figure}

The relativistic adiabatic index plays an important role in the study of the stability of relativistic stars. A star is assumed to remain charge-neutral, although it may become charged at various stages of its evolution. The gravitational force in such a star is counterbalanced by Coulomb repulsion, which keeps the stellar structure from collapsing into a point singularity. When the gravitational force is balanced by the combined effects of fermionic degenerate pressure and Coulomb repulsion, the star achieves stability. The adiabatic index Gamma for an anisotropic relativistic compact star is given as
\begin{equation*}
    \Gamma=\left(\frac{\rho+p_{r}}{p_{r}}\right)\frac{dp_{r}}{d\rho}.
\end{equation*}
For a particular stellar configuration, Bondi \cite{Bondi1964} analysed that a Newtonian isotropic sphere will be in a stable state if the adiabatic index $\Gamma>\frac{4}{3}$, and it gets changed for a relativistic anisotropic fluid sphere. For an anisotropic relativistic sphere, Chan et al. \cite{chan1993} proposed the following stability condition:
\begin{equation*}
    \Gamma>\frac{4}{3}+\left[\frac{r}{3}\frac{\rho_{0} p_{r_{0}}}{|{p_{r_{0}}^{'}}|}+\frac{4}{3}\frac{p_{t_{0}}-p_{r_{0}}}{r|p_{r_{0}}^{'}|}\right]_{max}
\end{equation*}
where $\rho_{0}$, $p_{r_{0}}$ and $p_{t_{0}}$ are the fluid’s initial density, radial pressure and tangential pressure, respectively. The above condition clearly shows that for a stable anisotropic configuration, the limit on the adiabatic index depends upon the type of anisotropy. In 1975, Heintzmann and Hillebrandt \cite{HH1975} showed that the adiabatic index $\Gamma>\frac{4}{3}$ at every interior point of a relativistic anisotropic compact object whose anisotropy factor is positive and increasing. According to Chandrasekhar \cite{C1964}, relativistic correction may result in instabilities within the compact star. To address this, Moustakidis (\cite{M2017} suggested a more stringent condition $\Gamma\ge\Gamma_{crit}$, where $\Gamma_{crit}=\frac{4}{3}+\frac{19}{21}u$ is the critical value of the adiabatic index with $u$ being the compactness parameter. We analysed here the nature of the adiabatic index in the radial direction and ensured that $\Gamma\ge\Gamma_{crit}$. Figure \ref{fig:Figure. 5} shows that $\Gamma>\frac{4}{3}$ in the anisotropic stellar configuration, indicating that our model meets stability criteria.

\subsection{Energy Conditions}
\begin{figure}
\centering
    \includegraphics[height=.25\textheight]{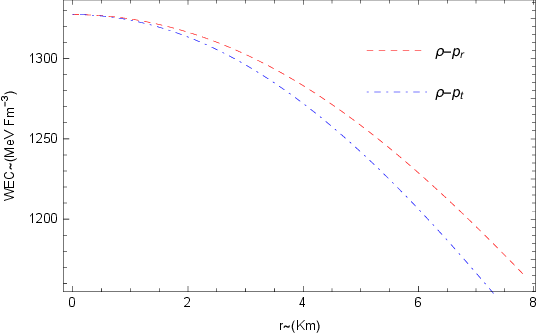}
    \caption{Weak Energy Condition against radial coordinate ($r$)}
    \label{fig:Figure. 6}
\end{figure}

\begin{figure}
\centering
    \includegraphics[height=.25\textheight]{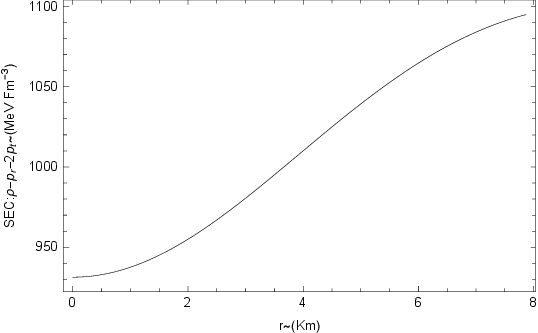}
    \caption{Strong Energy Condition ($\rho-p_{r}-2p_{t}$) against radial coordinate ($r$)}
    \label{fig:Figure. 7}
\end{figure}

Energy conditions play an important role for a compact star model in general relativity. It is necessary for the stellar composition to satisfy the energy conditions in order for it to be a physically acceptable model. These conditions include the null energy condition (NEC), the weak energy condition (WEC) and the strong energy condition (SEC). It is also significant for understanding the nature of the distribution of matter \cite{GV2003}. These conditions are defined as \cite{Leon1993}, \cite{Visser1995}.
\begin{equation} \label{NEC}
    \text{NEC: }\rho\ge0,
\end{equation}
\begin{equation}
    \text{WEC: }\rho-p_{r}\ge0, \rho-p_{t}\ge0,
\end{equation}
\begin{equation} \label{SEC}
    \text{SEC: }\rho-p_{r}-2p_{t}\ge0.
\end{equation}
To check all of the inequalities defined above, we used the graphical representations in Figures \ref{fig:Figure. 6} and \ref{fig:Figure. 7} to construct the profiles of \ref{NEC} - \ref{SEC}. Figures \ref{fig:Figure. 6} and \ref{fig:Figure. 7} illustrates that the aforementioned energy conditions are fulfilled throughout the interior of the stellar configuration.

\subsection{Electric Field}
\begin{figure}
\centering
    \includegraphics[height=.25\textheight]{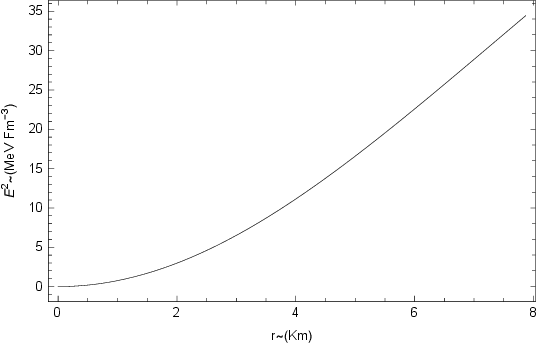}
    \caption{Square of Electric field intensity ($E^{2}$) against radial coordinate ($r$)}
    \label{fig:Figure. 8}
\end{figure}

A charged particle is referred to as a source of force in astrophysics. The forces might be gravitational, electromagnetic, or strong nuclear. We frequently identify the charged article by considering the balance between the number of protons and electrons in an atom. The magnitude of the charge at the center must be zero, and it reaches its peak at the boundary of the star \cite{MG2017}. Figure \ref{fig:Figure. 8} shows the square of the electric field $\left(E^{2}\right)$ plotted versus the radius of the star $\left(r\right)$. The Figure \ref{fig:Figure. 8} illustrates that the electric field is zero at the center and monotonically increasing.

\section{Conclusion}\label{sec6}
In this work, we have developed a new model of a charged anisotropic compact star in isotropic coordinates that is free from singularities. Several authors used an equation of state for stars, $p = p\left(\rho\right)$, to develop a physically plausible stellar model. Ngubelanga et al. \cite{N2015a} generated physically valid relativistic models of compact stars in isotropic coordinates using a linear equation of state. Ngubelanga et al. \cite{N2015b} employed the quadratic equation of state to determine the solution of an anisotropic distribution. Using the polytropic equation of state, Ngubelanga $\&$ Maharaj \cite{NM2015} obtained solutions for a relativistic star. Tell-Ortiz et al. \cite{TO2020} obtained the exact solution of Einstein's field equations using the modified generalised Chaplygin equation of state. Prasad et al. \cite{Prasad2021} developed a model for anisotropic compact stars, making use of the Chaplygin equation of state. Malaver \cite{M2013} used the modified Van der Waals equation of state to develop a physically plausible model. Since at ultra-high density, the equation of state may not be known, we developed a model where the equation of state is not precisely known. A noteworthy feature of our model is that we made no assumptions about any equation of state that relates the pressure and density of the stellar configuration. In this study, we generalised the model proposed by Ratanpal and Suthar \cite{Suthar2024} by incorporating charge into the fluid distribution. The singularity-free solution of Einstein-Maxwell field equations in isotropic coordinates has been obtained by assuming a physically reasonable metric potential together with a pressure anisotropy and electric field intensity. We investigated the physical validity of our model by analysing the particular model of pulsar \textbf{4U 1538-52}. A physical analysis reveals that our generated charged anisotropic model satisfies all of the physical characteristics of a realistic star, including the regularity of the gravitational potential at the origin, the positive definiteness of energy density, radial pressure, and tangential pressure at the origin, and the monotonically decreasing nature of the energy density, radial pressure, and tangential pressure. We also demonstrated that the model meets the energy conditions, which include null energy, weak energy, and strong energy. For analysing the stability of our model, we verified that the adiabatic index is greater than $\frac{4}{3}$ and represented graphically that our model meets the causality condition. It is difficult to find mathematical models of charged anisotropic fluid spheres that meet all physical constraints for stellar bodies such as compact stars. We have demonstrated that our solution can be utilized as a plausible model that can successfully explain the features of massive objects like neutron stars, quark stars, or other super-dense objects.

\backmatter

\bmhead{Acknowledgements}
The authors would like to thank the Inter-University Centre for Astronomy and Astrophysics (IUCAA) in Pune, India, for providing research facilities.


\bibliography{manuscript}

\end{document}